\documentclass{aa} 
\usepackage{graphics,amsmath,amssymb,rotating}

\begin{document}

\title{The Distance Scale of Planetary Nebulae}

\author{T. Bensby
 \and I. Lundstr\"om}

\offprints{T. Bensby, \\
\email{thomas@astro.lu.se}}

\institute{Lund Observatory, Box 43, S-22100 Lund, Sweden}

\date{Received / Accepted}

\abstract{
By collecting distances from the literature, a set of 73 planetary nebulae with mean distances 
of high accuracy is derived. This sample is used for recalibration of the mass-radius relationship, 
used by many statistical distance methods. An attempt to correct for a statistical peculiarity, where 
errors in the distances influences the mass--radius relationship by increasing its slope, has been made 
for the first time. Distances to PNe in the Galactic Bulge, derived by this new method as well as other 
statistical methods  from the last decade, are then used for the evaluation of these methods as distance 
indicators. In order of achieving a Bulge sample that is free from outliers we derive new criteria for 
Bulge membership. These criteria are much more stringent than those used hitherto, in the sense that they 
also discriminate against background objects. By splitting our Bulge sample in two, one with optically 
thick (small) PNe and one with optically thin (large) PNe, we find that our calibration is of higher 
accuracy than most other calibrations. Differences between the two subsamples, we believe, are due to the 
incompleteness of the Bulge sample, as well as the dominance of optical diameters in the ``thin'' sample 
and radio diameters in the ``thick'' sample. \newline
Our final conclusion is that statistical methods give distances that are at least as accurate as the ones
obtained from many individual methods. Also, the `long' distance scale of Galactic PNe is confirmed. 
}

\maketitle

\section{Introduction}

Planetary nebulae (PNe), being formed at the end of the asymptotic giant branch (AGB), constitute a short 
but important phase in the evolution of low and intermediate mass stars. Potentially, they can be used as 
probes of the distribution and kinematics of both their progenitors, i.e. AGB stars, and their end 
products, i.e. white dwarfs. The PNe show a strong concentration towards the Galactic Centre (GC). Among 
the $\sim1500$ known PNe, more than 25\% are located in the direction of the Galactic Bulge (from here on 
only referred to as `the Bulge'). The great majority of these are most probably situated within the Bulge,
and they are interesting as  targets both as individual objects and for statistical treatment. 

A major obstacle in most investigations of PNe is the lack of accurate distance determinations. A large 
collection of so called `statistical' distance determination methods have been proposed, each using only 
a few observational properties and using more or less well founded assumptions on the mean values of other
parameters needed, such as the mass of the ionized nebular envelope and the filling factor.

All of these methods are in some way related to the method by Shklovsky (\cite{shklovsky}), in which the 
main assumptions are that all PNe are optically thin spheres of constant density  and have the same 
nebular mass. Given the large variations in morphology and other observable parameters among PNe, it is 
not very likely that these assumptions are representative for most PNe. Usually, modifications to the 
original method involve changes of one or more of the assumptions above. An often used modification has 
been to replace the assumption of a constant ionized mass with a correlation of the nebular mass with 
observable quantities.

In a recent paper by Ciardullo et al.~(\cite{ciardullo}), a comparison between distances obtained through 
the method of identification of resolved binary companions of PNe nuclei and several statistical distance 
scales is made. Their results suggest that practically all statistical methods gives distances that are 
too large, except for older ones such as the method by Daub~(\cite{daub}). Newer  distance scales that in 
general are longer have therefore been put into question.

In this paper we will investigate the question of the accuracy of the statistical methods, and if they 
produce distances that are reliable enough. We will start by discussing the major statistical distance 
methods that have been proposed during the last decade, Sect.~\ref{sec:methods}, and point on the weaknesses 
these methods contain. Sect.~\ref{sec:standardpn} is then devoted to the problem of achieving a sample of 
PNe with individually determined distances that can be used for calibration of the mass-radius method, 
which is carried out in Sect.~\ref{sec:newmethod}. In Sect.~\ref{sec:bulge} we define new criteria for Bulge 
PNe that not only weed out foreground objects, but background objects as well. Sect.~\ref{sec:evaluation} 
then analyzes the validity of our method and the effects of our criteria for Bulge membership. Finally in 
Sect.~\ref{sec:conclusions} we give some concluding remarks.

\section{Previous distance scales} \label{sec:methods}

In the original Shklovsky method one assumes that all PNe can be approximated by spheres of fully ionized 
(or `density bounded') hydrogen with a constant mass and a uniform density. The expansion of the nebula is
then reflected by a synchronous decrease in the density of the emitting gas. If the nebula is optically 
thin at radio wavelengths, its distance, in parsecs, can be shown to be (see e.g. Milne \& Aller, 
\cite{milne})
\begin{equation}
      D = \left[\frac{2.58\cdot10^{20}M^{2}_{\rm ion}\ln(9980\,t^{3/2})}
          {\varepsilon\,t^{1/2}\,\theta^{3}\,S_{\rm 5\, GHZ}}\psi\right]^{1/5},
\label{eq:shklovsky1}
\end{equation}
where $\theta$ is the angular diameter  of the nebula in seconds of arc, $\varepsilon$ is the filling 
factor, $M_{\rm ion}$ is the ionized nebular mass in solar masses, $t=T_{\rm e}/10^{4}$ K is the electron 
temperature, and $S_{\rm 5\,GHz}$ is the free-free continuum radio flux at the 5 GHz frequency given in  
Jansky\footnote{1 Jy = $10^{-26}$ Wm$^{-2}$Hz$^{-1}$}. $\psi$ is a term involving the singly and doubly 
ionized helium contribution and is given by
\begin{equation}
      \psi=\left[\frac{(1+y+x''y)(1+y+2.7x''y)}{(1+4y)^{2}}\right],
\label{eq:shklovsky2}
\end{equation}
where $y$ is the number abundance ratio of helium to hydrogen ($=N_{\rm He}/N_{\rm H}$) and $x''$ is the 
fraction of doubly ionized helium atoms ($=N_{{\rm He}^{++}}/N_{\rm He}$). Adopting the following values 
for some of the parameters: $y=0.11$, $x''=0.5$ and $T_{\rm e} = 10\,000$~K, Eq.~(\ref{eq:shklovsky1}) 
reduces to
\begin{equation}
      D=17\,580\,M^{2/5}_{\rm ion}\varepsilon^{-1/5}
        \theta^{-3/5}S_{\rm 5\, GHz}^{-1/5}.
\label{eq:shklovsky3}
\end{equation}

Usually, a constant value (normally $\sim0.5-0.6$) has been adopted for the filling factor. In order to 
avoid the assumption of a constant mass in the equation above, different authors have tried to find 
relations between either the masses, the surface brightness temperatures or the optical thicknesses and 
the radii of the nebulae. No matter what, in the end all distance scales end up with distances that are 
dependent of only two parameters: the 6 cm radio flux and the angular size, as measured from either 
optical or radio wavelengths. During the last decade several such distance scales have been proposed, 
notably Cahn, Kaler and Stanghellini~(\cite{cahn}) (hereafter CKS), Van de Steene and 
Zijlstra~(\cite{vandesteene}) (VdSZ), Zhang~(\cite{zhang2}) (Z95), and Schneider and 
Buckley~(\cite{schneider}) (SB96).

\begin{description}
 \item[\underline{\textbf{CKS:}}]
    Daub~(\cite{daub}) empirically related the nebular ionized mass to a thickness parameter 
    $\mathcal{T}$ defined by
    \begin{equation}
          \mathcal{T}=\log\left(\frac{\theta^{2}}{S_{\rm 5\,GHz}}\right).
    \end{equation}
    CKS then recalibrated Daub's distance scale using a selection of 19 PNe with well-known  distances. 
    Their new relationship was split in two, one for optically thick PNe ($\mathcal{T}<3.13$) and a 
    constant value for optically thin PNe ($\mathcal{T}>3.13$), resulting in distances given by
    \begin{equation}
          \!\!\!\log D =  \left\{ \begin{array}{ll}
          2.71+0.2\log\theta-0.6\log S, & \textrm{ $\mathcal{T} < 3.13$} \\
          3.96-0.6\log\theta-0.2\log S, & \textrm{ $\mathcal{T} > 3.13$}. \\
          \end{array} \right.
    \label{eq:cks}
    \end{equation}
    For the distance determination CKS used radio fluxes and optical diameters except for nebulae with 
    very small angular sizes, where radio diameters obtained with VLA were used. 

    The main source of uncertainty is the small number of PNe used in the calibration. The two different 
    relations for thin and thick nebulae are also not very well determined. \newline

 \item[\underline{\textbf{VdSZ:}}]
    This method is based on a relationship between the distance-independent radio continuum brightness 
    temperature given by  
    \begin{equation}
          T_{\rm b}=73200\frac{S_{\rm 5\,GHz}}{\theta^{2}}
    \end{equation}
    and the distance-dependent radius of the nebula, calibrated by a sample of 131 PNe in the Bulge, 
    assuming they all lie at a distance of 8.0 kpc. The criteria used for Bulge membership were: 
    $|l|<10\degr$, $|b|<10\degr$, $\theta<20\arcsec$, and $S_{\rm 5\,GHz}<100$ mJy. Distances obtained by 
    this method are given by
    \begin{equation}
          \log D=3.40-0.3\log\theta-0.35\log S_{\rm 5\,GHz}.
    \label{eq:vdsz}
    \end{equation}
    A worry is the way calibration nebulae were selected. Foreground nebulae are reasonably well weeded 
    out by the selection criteria used, but no attempts have been made to sort out background objects from
    the Bulge sample. \newline 

 \item[\underline{\textbf{Z95:}}]
    This method uses two correlations: one between the ionized mass and the radius, and one between the 
    radio continuum surface brightness temperature and the radius. The distance scale proposed is an 
    average of the two distances obtained using the correlations, and is given by
    \begin{equation}
          \log D_{\rm f}=3.39-0.27\log\theta-0.36\log S_{\rm 5\,GHz}.
    \label{eq:z95}
    \end{equation}
    The correlations were calibrated by 132 PNe with individual distances from Zhang~(\cite{zhang}).

    One could argue that a possible weakness of this relation is that it is based on only one particular 
    method for determining individual distances.  \newline

 \item[\underline{\textbf{SB96:}}]
    This method is based on a relationship between the PN radius and the radio surface brightness. The 
    nebular surface brightness, in Jy/arcsec$^{2}$, is given by
    \begin{equation}
          I=\frac{4\,S_{\rm 5\,GHz}}{\pi\theta^{2}},
    \end{equation}
    which is a distance-independent quantity. The best-fit relationship between the two parameters 
    results in distances given by
    \begin{equation}
          \begin{split}
          \!\!\log D=3.37-\log\theta-0.026(\log I)^{2}-0.46\log I.
          \end{split}
    \label{eq:sb96}
    \end{equation}
    This method differs from the previous ones in the sense that a second order polynomial is used in the 
    calibration. To a large extent their calibrating sample consisted of Bulge PNe for which standard 
    selection criteria (same as VdSZ) were used.

\end{description}

A first concern is the accuracy of the angular diameters used. Stasi\'nska et al.~(\cite{stasinska}) and 
Pottasch and Zijlstra~(\cite{pottasch4}) take opposite views regarding the accuracy of optical versus 
radio measurements of the angular diameters. SB96 also comments that ``the situation regarding the angular
measurements of PNs is in a sorry state''.

Secondly, the selection of a calibrating sample of PNe is not trivial. Usually, nebulae with ``good'' 
individual distance determinations are chosen, and such samples often include a large fraction of nearby 
nebulae. The definition of what constitutes a good distance determination is far from obvious. 
Alternatively, Bulge nebulae are used and in this case the selection of true members can be troublesome, 
as well as large errors in the observed parameters, notably the angular sizes.

Finally, there seem to exist a statistical peculiarity in the mass--radius relation that hitherto 
not has been payed attention to. Errors in the distances have the inevitable effect of increasing the 
slope of the observed distribution (see Sect.~\ref{sec:correction}).        

For these reasons we found it worthwhile to reanalyze the problem, paying special attention to the 
problems discussed above.

\section{The calibration sample} \label{sec:standardpn}

\subsection{Individual distances}

There are many different non-statistical methods for distance determinations of PNe or their central 
stars, e.g. from identification of a resolved binary companion, from expansion velocities, determination 
of surface gravities and luminosities etc. For a few PNe trigonometric parallaxes from ground-based 
measurements or from the Hipparcos mission are available. We have conducted a search through the 
literature for PNe with individually determined distances. 
It is important that the distance estimates are independent. Sometimes this might be an awkward 
thing to objectively decide upon. For example, the extinction maps produced by Lucke~(\cite{lucke}) have 
been used by many authors in order to estimate the amount of reddening towards a nebula. Should these 
distance estimates be treated as independent or not? Since the methods for interpolating and estimating 
the extinction differs between different authors, often with widely varying results, we believe that it 
is all right to do so.
 
A total of 285 PNe were found from papers spanning three decades in publication dates. For a PN to be 
qualified as a potential ``standard'' PN we demanded that it should have at least three
independent individual distance determinations. This leaves us with 105 PNe. For these PNe we 
calculated a mean distance ($D$) and the error of this mean ($\Delta D$) (computed as the standard 
deviation divided by the square root of the number of individual distance determinations). All PNe with 
$\Delta D / D > 0.2$ were then rejected, leaving 41 PNe to work with. We call these selected PNe 
`sample A' and they are listed in Table~\ref{tab:sample_A} in the Appendix.

In order to increase the size of our sample we subsequently looked at those PNe with only two individual 
distance determinations. We selected those that had $\Delta D / D < 0.2$ ($\Delta D$ is in this case just 
the difference between the two distances divided by 2). We call these 32 PNe `sample B', and they are 
listed in Table~\ref{tab:sample_B} in the Appendix. A majority of these PNe comes from the paper of 
Zhang~(\cite{zhang}), and as can be seen they are located at larger distances than the PNe in sample A. 
This also means that the relative errors in observable quantities, especially the angular diameters, 
generally are larger than for sample A. In the subsequent analysis, the nebulae in this second sample are 
given the weight 0.5 compared to those in sample A.

\subsection{Radio fluxes}

Radio fluxes for PNe have been obtained both through single dish and interferometric observations. 
Zijlstra et al.~(\cite{zijlstra4}) noted a systematic difference between their VLA observations at 5 GHz 
and the corresponding measurements in the Parkes survey (Milne and Aller,~\cite{milne3}), which they 
suggested to be primarily due to contamination from nearby sources in the single dish observations. For 
strong sources the difference was attributed to extended halos, which could be missed in the VLA 
measurements due to lack of sufficiently short baselines.

However, Tylenda et al.~(\cite{tylenda}) argue, through a comparison of optical and radio determinations 
of extinction constants, that some of the VLA fluxes, in particular those of 
Gathier et al.~(\cite{gathier4}), Pottasch et al.~(\cite{pottasch8}), and 
Zijlstra et al.~(\cite{zijlstra4}) seem to be systematically too low. Pottasch and 
Zijlstra~(\cite{pottasch9}) subsequently remeasured a subsample of these sources at lower resolution and 
found that for faint ($<10$ mJy) sources the old VLA fluxes were sometimes too low, but no discrepancies 
were found for brighter sources.

Most of the PNe observed at 5 GHz by Milne and Aller (\cite{milne}) and Milne (\cite{milne4}) were also 
observed at 14.7 GHz (Milne and Aller,~\cite{milne3}). They concluded that most PNe were optically thin 
at both frequencies. The smaller 14.7 GHz beam will make these observations less susceptible to 
contamination. In this investigation we have used the mean flux densities of the Parkes survey at 5 and/or
14.7 GHz and different VLA observations. The 14.7 GHz observations were converted to 5 GHz using the 
relation $S_{\nu}\propto\nu^{-0.1}$, i.e. assuming the sources to be optically thin at these wavelengths. 
References to the observations used are listed in Table~\ref{tab:sample_C} in the Appendix.

The differences between the converted 14.7 GHz fluxes and the observed 5 GHz fluxes were generally found 
to be small ($<10\%$ of the mean flux) except for a few faint sources ($<10$ mJy), indicating that almost 
all nebulae are in fact optically thin at both frequencies. There are, however, two nebulae that are 
optically thick at both 5 and 14.7 GHz, viz. Vy 2-2 and Hb 12 (Purton et al., \cite{purton}; Seaquist 
and Davis, \cite{seaquist}; Aaquist and Kwok, \cite{aakw2}). For these nebulae, we have calculated the 
flux densities expected at 5 GHz if the nebulae had been optically thin from the observations at higher 
frequencies presented by these authors.

\subsection{Angular diameters}

As stated in Sect.~\ref{sec:methods}, angular diameters are, for numerous reasons, not always easy to 
determine. Firstly, only very few PNe are truly spherical, and some have indeed very irregular shapes. 
Secondly, many nebulae show extended low intensity halos and/or multiple shell structures and several 
largely different values for the diameter of a particular nebula are often found in the literature. 
Thirdly, for small nebulae seeing, tracking errors etc. limit the accuracy that can be obtained. Moreover,
the angular size is depending on the wavelength region observed, e.g. [O{\sc iii}] diameters are usually 
smaller than those measured in H$\alpha$. In radio determinations of angular diameters, several different 
methods are used (see e.g. Zijlstra et al., \cite{zijlstra4}), and as will be further discussed in 
Sect.~\ref{sec:bulge}, sometimes optical and radio measurements show large discrepancies.

Ideally, the angular sizes of all PNe  should be measured in a consistent way. In radio observations, one 
often use the 10\% contour level of the peak flux density. This method, however, can only be used for 
bright and well resolved nebulae. Moreover, a rather large fraction of the flux could fall outside this 
limit, and a lower level might be preferred. However, inspection of published data and images from the HST
data archive show that the flux usually drops very fast between contour levels in the interval 15 to 5\%, 
so the choice of level is not very critical.

In this investigation we have in most cases adopted angular diameters for the calibration nebulae from 
the Strasbourg-ESO Catalogue (SECAT), Acker et al.~(\cite{acker3}), (rejecting all PNe that have 
uncertainty flags and/or upper/lower limit values), after checking with sky survey plates and in many 
cases with images from the HST Data archive. These checks have in some cases led to significant revisions 
of the diameters. On HST images, diameters have been determined from the contour at $\sim$5\%  of peak 
intensity. For elongated nebulae the mean of the major and minor axis has been used. For references to the
sources of the adopted diameters, see Table~\ref{tab:sample_C} in the Appendix.

\section{Calibration of the distance method} \label{sec:newmethod}

\begin{figure*}
 \resizebox{\hsize}{!}{\includegraphics{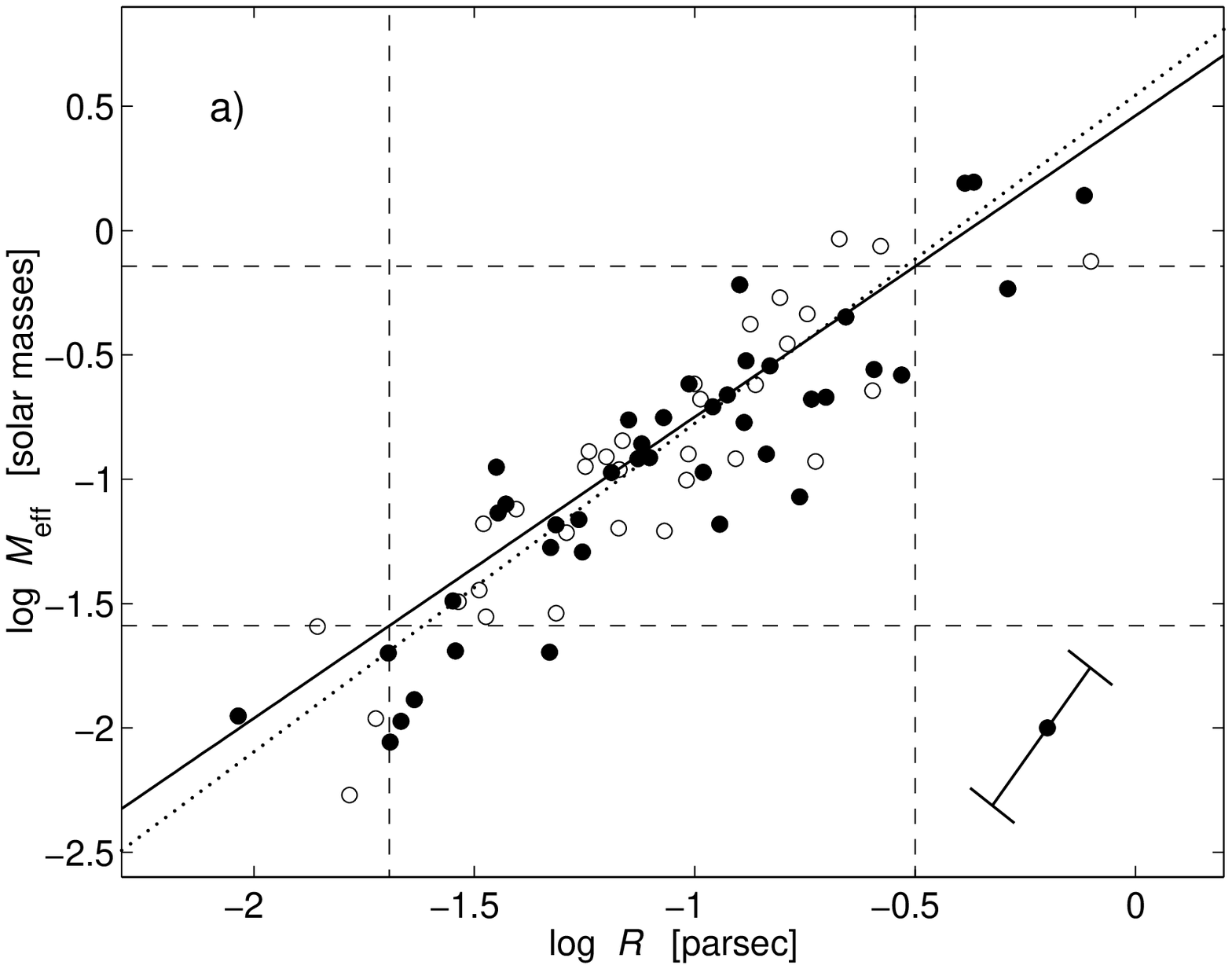}
                       \includegraphics{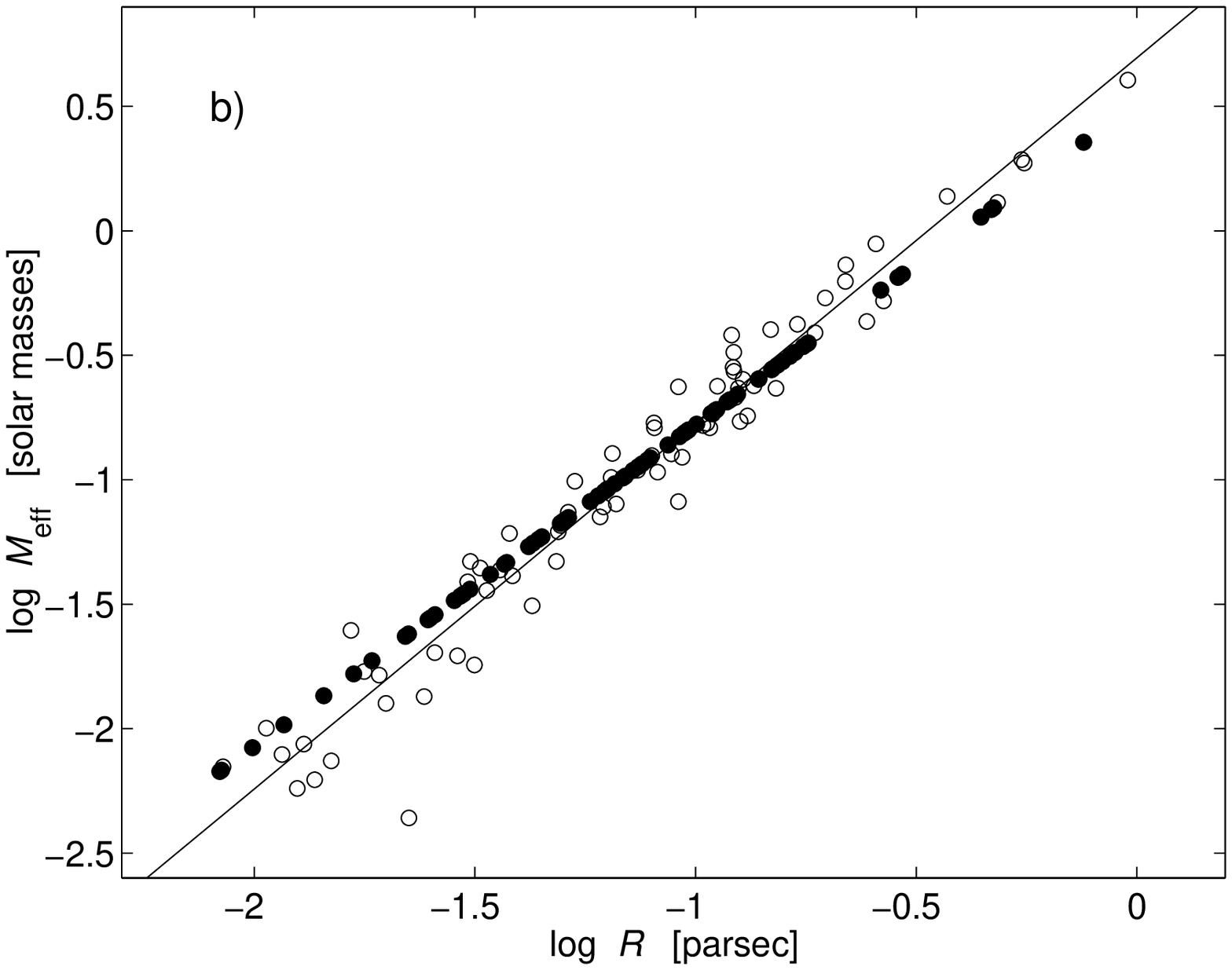}}
 \caption{{\bf a)} 
                  Effective masses versus radii for the calibration samples. Filled circles represent 
                  those 41 PNe that have three or more independent distance determinations, open circles 
                  those with just two. Two regression lines are also plotted: the original weighted one 
                  (dotted line, see Eq.~(\ref{eq:constants})) and the one corrected for uncertainties in 
                  the adopted distances (solid line, see Eq.~(\ref{eq:constants3})). The error bar 
                  indicates how a data point is affected by an uncertainty of 25\% in the adopted 
                  distances. Dashed lines are explained in Sect.~\ref{sec:criteria}. \newline
          {\bf Fig. 1. b)} 
                  A simulated sequence of PNe lying perfectly along the mass-radius relationship (filled 
                  circles). Open circles show how the distribution is steepened when random errors of 30\%
                  in the distances are introduced. A regression line is also plotted.}
\label{fig:massaradie}
\end{figure*}

\subsection{Mass - radius relationship} \label{sec:massradius}

From Eq.~(\ref{eq:shklovsky3}) we define an `effective' mass as 
\begin{equation}
      M_{\rm eff}\equiv\frac{M_{\rm ion}}{\varepsilon^{1/2}}=2.44\cdot
      10^{-11}D^{5/2}\cdot \theta^{3/2} \cdot S_{\rm 5\, GHz}^{1/2}.
\label{eq:shklovsky4}
\end{equation}
For an ideal PNe ($\varepsilon=1$), this effective mass is identical to the usual ionized mass. The 
filling factor only corrects for (by increasing the mass) that the emitting gas volume is smaller than the
one that is obtained by measuring the angular size (which otherwise would result in a too low density).

The PNe radii, in parsecs, are given by
\begin{equation}
      R = \tan \left(\frac{\theta/2}{206265}\right)\cdot D 
      \simeq \frac{\theta/2}{206265}\cdot D.
 \label{eq:pnsize}
\end{equation}

Fig.~\ref{fig:massaradie}a shows the effective masses versus the sizes of the PNe in the two samples 
(A and B). There are no doubts that the masses keep rising for all radii.

A linear relationship,
\begin{equation}
      \log M_{\rm eff} = \alpha \cdot \log R + \beta,
 \label{eq:regression1}
\end{equation}
is then fitted by weighted least square regression. The constants $\alpha$ and $\beta$ are derived by 
minimizing the chi-square merit function (see for instance Press et al.~(\cite{press}))
\begin{equation}
   \chi^{2}(\alpha, \beta) = \sum_{i=1}^{N}\left(\frac{(\log M_{\rm eff})_{i}-\alpha-
   \beta \cdot (\log R)_{i}}{\sigma_{i}}\right)^{2},
\end{equation}
where $\sigma_{i}$ is the uncertainty associated with each $\log M_{\rm eff}$, i.e. 2.5 times the 
uncertainty in the distance. Distance uncertainties in sample B, that are given less weight compared to 
sample A, are also multiplied by a factor two. An approximation here is that we assume the radii to be 
known exactly (which is not true since errors in the distances translates to the radii as well, although 
not as much as for the masses). The resulting regression line is given by  
\begin{equation}
      \alpha=1.32\pm0.04, \,\,\,\,\,\,\,\,
      \beta=0.55\pm0.05.
 \label{eq:constants}
\end{equation}
and is plotted as a dotted line in Fig.~\ref{fig:massaradie}a.

The correlation coefficient of the fit is 0.90 and the standard deviation of the residuals of 
$\log M_{\rm eff}$ is 0.25. We also tried out other fits to the data: non-weighted fits and second order 
fits, of which none gave any improved solution.

\subsection{Corrections to the fit} \label{sec:correction}
 
How do errors in the adopted distances of our calibration sample affect the mass--radius 
relationship? The error bar in Fig.~\ref{fig:massaradie}a shows the displacement a data point will 
undergo for an error of 25\% . As can be seen, this `error vector' 
($\log M_{\rm eff}$ is proportional to $2.5 \log D$ and therefore also to $2.5\log R$) 
is roughly aligned with the derived 
regression line, only slightly steeper. A consequence of this, that was pointed out by our referee 
Dr. A. Zijlstra, is that the observed distribution is steeper than it would be if the distances were error
free. Our derived regression coefficients are then too large.

To correct for this effect we will simulate a sequence of PNe lying along our derived 
relationship. By introducing random errors in the distances of these PNe, and fitting a new linear 
relationship we will be able to see how large the steepening is. The effect will of course be more 
prominent the larger the errors in the distances are. From the fit above we saw that the residuals are 
about 0.25 (in $\log M_{\rm eff}$, i.e. vertically aligned). If the spread's only source is inaccurate 
distances, this corresponds to a shift of about 0.57 along the error vector back to to regression line. 
Then an uncertainty of about 70\% in the distances of our calibration sample is implied. However, there 
are errors coupled to the angular sizes and the radio fluxes as well. Then we have different assumptions 
about the electron temperature, degree of ionization, the geometry of the nebulae etc. Also, the reasoning
above is based on the fact that there exists an absolute relationship between the masses and the radii for
the nebulae. If not, there is also a kind of intrinsic ``cosmic'' scatter included in the plot. Therefore,
maybe we can assume that the adopted distances are responsible for about 50\% of the spread, i.e. 
$\sim 0.12$ in the residuals of $\log M_{\rm eff}$. Although this assumption is somewhat precarious, it 
implies an accuracy of about 30\% in the adopted distances.

Fig.~\ref{fig:massaradie}b shows a sample of 73 PNe (filled circles) lying perfectly along a 
mass-radius relationship. Empty circles show how random errors in the distances (Gaussian distribution 
with a $\sigma$ of 30\%) affects the distribution. From the fitting of a regression line to this 
distribution we see that the slope is increased by 9\% and the constant term by 18\% (computed as the 
average from hundreds of simulations, the one shown in the figure is just one representive case).

Compensating the coefficients in Eq.~(\ref{eq:constants}) for this effect (i.e. dividing $\alpha$ 
and $\beta$ by 1.09 and 1.18 respectively) gives the final relationship (shown as the solid line in 
Fig.~\ref{fig:massaradie}a)
\begin{equation}
      \log M_{\rm eff} = 1.21 \cdot \log R + 0.46.
\label{eq:constants3}
\end{equation} 

One striking thing that can be seen in Fig.~\ref{fig:massaradie}b is how the down-dip on the 
lower left, seen in the original data, is reproduced by introducing errors to the distances. This 
indicates that it might be a kind of artifact and warrants the use of a linear mass-radius relationship.

\subsection{Resulting method}

By invoking the effective mass-radius relationship from Eq.~(\ref{eq:regression1}) in 
Eq.~(\ref{eq:shklovsky4}), and making use of Eq.~(\ref{eq:pnsize}) it is easy to show that the distances 
(in parsecs) are given by
\begin{equation}
      \begin{split}
      \log D = \frac{5.61\alpha-\beta-10.61}{\alpha-2.5}+\frac{1.5-\alpha}
               {\alpha-2.5}\log\theta  \\
               +\frac{0.5}{\alpha-2.5}\log S_{\rm 5\,GHz}, 
      \end{split}
\label{eq:new}
\end{equation}
which through the result of the fit, Eq.~(\ref{eq:constants3}), becomes
\begin{equation}
      \log D = 3.31 - 0.22\,\log\theta - 0.39\,\log S_{\rm 5\,GHz}.
 \label{eq:new2}
\end{equation}
This distance relation is really not that different from the ones by VdSZ and Z95. A trend is 
quite obvious though; our method is less dependent on the angular size, but slightly more on the radio 
flux, and the zero point is lower. The differences are slightly smaller compared to Z95 than compared to 
VdSZ. Consequences due to  these differences are analyzed in Sect.~\ref{sec:newdistances}.

\section{Bulge PNe} \label{sec:bulge}

\subsection{Selection criteria} \label{sec:criteria}

For a PN to qualify as a member of the Galactic Bulge, it must have certain properties. First of 
all it must be located in the right direction. The most widely used criteria for the galactic coordinates 
are $|l|<10\degr$ and $|b|<10\degr$. At a distance of 8 kpc it corresponds to a Bulge radius of about 1.5 
kpc. Further limitations on the nebulae are set upon their angular diameters ($\theta<20\arcsec$) and 
their 6 cm radio continuum flux ($S_{\rm 5\,GHz}<100$ mJy). These four criteria are what have been used 
hitherto. What they do is to weed out foreground objects, and do so very effectively. Further criteria, in
order to also sort out background objects, have been neglected, resulting in Bulge samples that are 
polluted by background PNe. Since the GBPNe are often used in the evaluation of new distance methods, and 
even for calibration purposes, it is very important to get a sample of Bulge PNe, that are likely to be 
true members, and that it is as pure from outliers as it can possibly be. To achieve this we will rework 
the criteria that a PN must fulfill to be treated as a member of the Bulge.

The distance to the GC is an important parameter when estimating what a typical PN in the Bulge looks 
like. Reid~(\cite{reid}) gives 8.0 kpc as the best distance, which also is the value that more recently 
published values cluster around, see Binney and Merrifield~(\cite{binney}). It is a weighted mean of a 
variety of methods and should be a reliable estimate. 

We further assume a Bulge radius of 1.5 kpc, i.e the maximum distance a Bulge nebula will have is 9.5 kpc,
and the minimum is 6.5 kpc.

\subsubsection{Size limits} \label{sec:sizelimit}

Fig.~\ref{fig:massaradie}a shows the range in radii for our PNe used in the calibration. Since no 
attempts have been made in trying to put upper or lower limits to these radii, we believe that the range 
in size that those PNe represent is in a way representive for PNe in general. Upper and a lower limit on 
the sizes of Bulge nebulae can then be set. To do so we reject the lower and upper 8\% 
($\sim1.5$ standard deviation from the mean) of the PNe in the diagram, leading to $-1.69<\log R < -0.50$.
These limits are plotted as the two dashed vertical lines in Fig.~\ref{fig:massaradie}a. The allowed 
angular size range for Bulge PNe then become, through Eq.~(\ref{eq:pnsize}),
\begin{equation}
      0.9\arcsec<\theta<20\arcsec,
\label{eq:crit1}
\end{equation}
where maximum size is associated with minimum distance and maximum radius (i.e. $D=6500$ pc and 
$\log R=-0.50$) and vice versa (i.e. $D=9500$ pc and $\log R=-1.69$) for minimum size.

\subsubsection{Flux limits} \label{sec:fluxlimits}

By inserting the limits of $\log R$ into our mass--radius relationship, Eq.~(\ref{eq:constants3}), we may 
also set lower and upper limits to the effective mass. This leads to $-1.59<\log M_{\rm eff} < -0.14$ 
(plotted as dashed horizontal lines in Fig.~\ref{fig:massaradie}a). The reason why we use the 
mass--radius relation, instead of only reading these limits in Fig.~\ref{fig:massaradie}a, is that the 
masses are more sensitive to errors in the derived distances than the radii 
($\log M_{\rm eff} \propto 2.5\log D$ and $\log R \propto \log D$).

Maximum radio flux is associated with minimum size (and therefore also minimum mass) and minimum distance 
(i.e. $\log R \sim -1.69$, $\log M_{\rm eff} \sim -1.59$, and $D\sim 6.5$ kpc), and vice versa 
(i.e. $\log R \sim -0.50$, $\log M_{\rm eff} \sim -0.14$, and $D\sim 9.5$ kpc) for minimum radio flux. 
Using again Eq.~(\ref{eq:shklovsky4}) and solving for the radio flux gives
\begin{equation}
       4.2\,\,{\rm mJy} < S_{\rm 5\,GHz} < 45\,\,{\rm mJy}. 
\label{eq:crit2}
\end{equation}

In Fig.~\ref{fig:criteria} we plot the new criteria (besides the coordinate critoria), 
Eqs.~(\ref{eq:crit1}) and (\ref{eq:crit2}), for Bulge membership. They form a rectangle that restricts 
the values of the angular diameter and the radio flux for a PN to be treated as a Bulge nebula. These 
criteria will most likely also weed out a few PNe that lie in the Bulge due to their tightness. In order 
to get the sample as clean as possible this must be endured, at least as long as outliers are effectively 
deleted.

\subsection{Optical or radio diameters?}

The number of PNe fulfilling the coordinate criteria in the SECAT with optical diameters is 245, and with 
radio diameters 136. Although there is a small overlap, the radio diameters peak at considerably smaller 
values than the optical ones.

The discrepancies, as was also pointed out in Sect.~\ref{sec:methods}, between radio and optical measurements
of the angular diameter, makes it hard to decide which of the two sources to use for PNe in the Bulge. It 
is especially for diameters below $5\arcsec$ where the disparities are. Due to the large distances it is 
not possible to proceed as for our samples A and B, where sky survey plates and in many cases images from 
the HST Data archive were used to check on the diameters.

To see if there are differences arising from using either optical or radio measurements of the angular 
diameter, a set of 61 PNe, taken from the SECAT, whose radio fluxes and angular diameters (both optical 
and radio) fulfill our new Bulge criteria was analyzed. Distances obtained  (using our new method as well 
as the other methods discussed in Sect.~\ref{sec:methods}) with optical diameters appear to be marginally 
shorter than those obtained using radio diameters. And except that radio distances seem to be more nicely 
distributed and slightly closer to the assumed 8 kpc distance to the GC, there are really no major 
differences. This makes it very difficult to judge which of the two sources for the angular sizes of PNe 
that is the most reliable. To minimize the risk of choosing the most erroneous value when they differ a 
lot, a mean value of the optical and the radio values might be to prefer. Assume that they differ by as 
much as 50\% compared to the mean value of the two. If this difference is due to an error in just one of 
the values, and if the wrong one is chosen, it will lead to an error of $\sim10$\% in the derived 
distance, see Eq.~(\ref{eq:new2}). By instead using the mean value, the error in the derived distances 
due to erroneous diameters is reduced to a maximum of $\sim5$\% for a 50\% disparity.

\subsection{A sample of Bulge nebulae} \label{sec:bulgesample}

\begin{figure}
      \resizebox{\hsize}{!}{\includegraphics{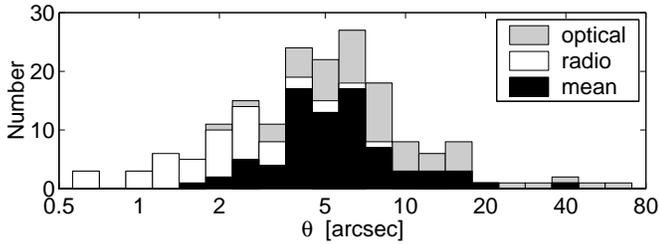}}
      \caption{Angular diameters for the 174 PNe that fulfill the coordinate criteria for Bulge 
               membership. White parts represent PNe with radio diameters, grey PNe with optical 
               diameters, and black PNe where a mean value of the radio and optical has been used.}
\label{fig:size_and_flux}
\end{figure}
\begin{figure}
      \resizebox{\hsize}{!}{\includegraphics{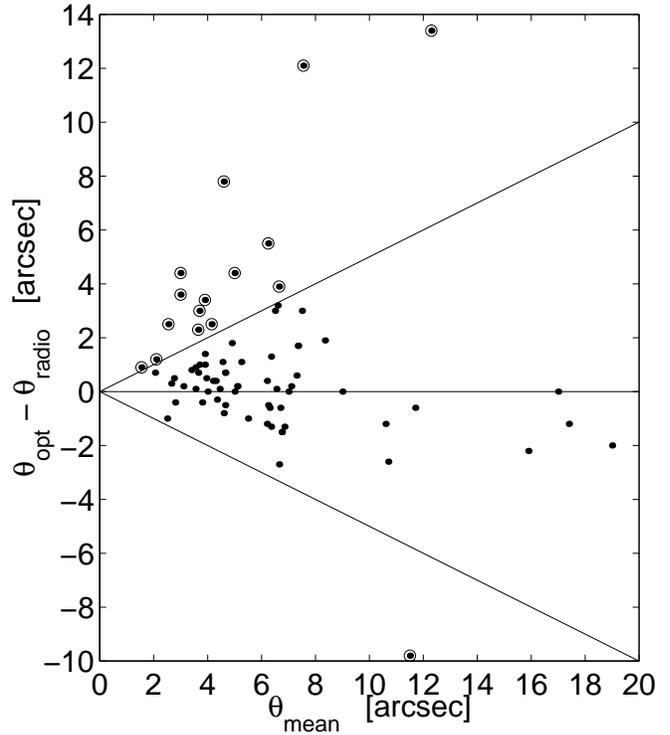}}
      \caption{In the sample of those PNe that fulfill the coordinate criteria there are 77 for which a 
               mean value of the optical and radio diameters has been used. Here we plot the optical 
               values minus the radio values as a function of the mean. Encircled PNe are the ones that
               differs by more than 50\%, in all 16.}
 \label{fig:sizediff}
\end{figure}
\begin{table}
 \centering
 \caption[]{For those 10 PNe that high-resolution images are available we list here their radio and 
            optical diameters, taken from the SECAT, as well as the diameters that we adopted after 
            inspection of the images. The references are Dopita et al.,~\cite{dopita} (D90); 
            Bedding \& Zijlstra,~\cite{bedding} (BZ); and the HST Archive. The diameters from BZ have been
            reduced by $\sim 10$\%, see discussion in the text. In all cases the adopted value fall in 
            between the optical and the radio, indicating that using the mean value is an acceptable 
            method.}
 \begin{tabular}{crrrl}
 \hline\noalign{\smallskip}
     PN G      & Optical & Radio & Adopted & Ref.  \\
 \noalign{\smallskip}
 \hline\noalign{\smallskip}
 \noalign{\smallskip}

  002.1$-$04.2 & 4.8~~  & 1.2~~  & 2.0~~~  & HST, D90  \\
  003.5$-$04.6 &13.6~~  & 1.5~~  & 9.0~~~  & HST  \\
  004.6$+$06.0 & 8.6~~  & 4.7~~  & 4.7~~~  & BZ  \\
  005.0$+$04.4 & 5.2~~  & 0.8~~  & 1.0~~~  & BZ  \\
  350.9$+$04.4 & 5.6~~  & 2.2~~  & 2.8~~~  & HST \\
  355.4$-$02.4 & 7.2~~  & 2.8~~  & 4.7~~~  & BZ  \\
  355.7$-$03.5 & 2.0~~  & 1.1~~  & 1.7~~~  & BZ  \\
  357.4$-$03.2 & 5.4~~  & 2.2~~  & 2.3~~~  & BZ  \\
  358.2$+$04.2 & 5.4~~  & 2.9~~  & 2.9~~~  & BZ  \\
  359.9$-$04.5 & 4.8~~  & 2.5~~  & 3.1~~~  & D90 \\

 \hline
 \end{tabular}
\label{tab:highres}
\end{table}

To maximize the number of PNe, both optical and radio values for the angular sizes will be used, and when 
both are available a mean value of the two. For PNe that only have the 2 cm radio flux measured, we 
convert it into a corresponding 6 cm flux through the relationship $S_{\nu}\propto\nu^{-0.1}$. We also 
reject those PNe that have uncertainty flags on any of these parameters, or if the values are marked as 
upper/lower limits. In all this leaves 174 PNe towards the Bulge that fulfill the coordinate criteria.  
Fig.~\ref{fig:size_and_flux} shows the angular size distribution for these PNe. Of these there are 77 
for which a mean diameter has been used. Fig.~\ref{fig:sizediff} shows the difference between the 
optical and the radio values as a function of the mean value. There are 16 that differs by more than 50\%,
i.e. $\sim20$\% of the PNe that have both radio and optical values.

For 10 of these 16 objects high-resolution optical images exist (Dopita et al.,~\cite{dopita}; 
Bedding and Zijlstra,~\cite{bedding} and HST image archive). In all cases, diameters determined from these
images fall between the optical and the radio diameters given in SECAT, see Table~\ref{tab:highres}, 
indicating that the use of mean values is an acceptable method. For the objects discussed here, we have 
however used the diameters determined from the high-resolution images. Of the remaining six objects, three
fail to meet our flux criteria (Sect.~\ref{sec:fluxlimits}), leaving only three objects with strongly 
deviating optical and radio diameters, which we choose to reject from our sample.

For the PNe with largely deviating diameters, the optical one is always larger than the one 
measured in radio. van Hoof (\cite{vanhoof}) discusses this in detail, and it seems as if the effect is 
real. This is due to the fact that a conversion factor has to be applied in order to achieve a true 
diameter from an observation at limited resolution. van Hoof shows that this conversion factor is about 
10\% larger for radio emission than for H$_{\alpha}$ emission. As Bedding \& Zijlstra (\cite{bedding}) 
used the radio conversion factors for the optical images, they found roughly 10\% larger diameters in the 
H$_{\alpha}$ images. In Table~\ref{tab:highres} we have reduced the `BZ' diameters by 10\%. Four of the 
objects then coincide with the radio diameters listed in SECAT.

\begin{figure}
      \resizebox{\hsize}{!}{\includegraphics{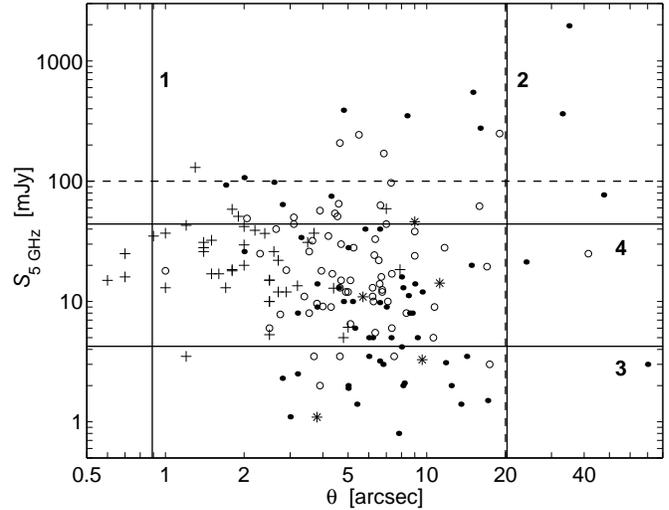}}
      \caption{Radio fluxes versus angular diameters for 174 PNe towards the Bulge located within 
               $|l|<10\degr$ and $|b|<10\degr$. The different markers are: 
               `$\bullet$' -- optical diameters; 
               `$\circ$' -- mean of optical and radio diameters;
               `$+$' -- radio diameters; 
               `$*$' -- 2 cm fluxes converted to 6 cm fluxes and optical diameters. 
               The solid lines marked $1-4$ represent the limits for our new Bulge criteria 
               (see Sect.~\ref{sec:criteria}), where 109 are approved. For comparison we also plot the old 
               criteria, i.e. $S_{\rm 5\,GHz}<100$ mJy and $\theta<20\arcsec$ (dashed lines). }
\label{fig:criteria}
\end{figure}

Fig.~\ref{fig:criteria} shows how these 174 PNe are distributed in the 
`$\log S_{\rm 5\,GHz} - \log\theta$' plane. In the figure we also plot our new selection criteria (solid 
lines marked $1-4$). Those 109 PNe that lie within the rectangle are the ones that fulfill the criteria 
for Bulge membership. For comparison we plot the old criteria (dashed lines). We see that there are PNe 
with low radio fluxes and/or small angular sizes that fail to make the new demands for membership. These 
PNe are supposed to be background objects. Also the upper limits are tighter and weed out more 
(foreground) objects than before. 

\section{Evaluation} \label{sec:evaluation}

\subsection{Local PNe -- Bulge PNe}

\begin{figure}
      \resizebox{\hsize}{!}{\includegraphics{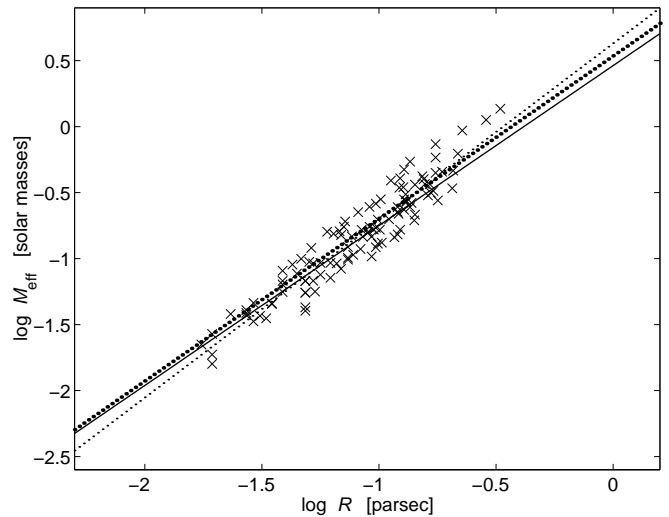}}
      \caption{Mass--radius distribution for the Bulge sample. Thin dotted line is a regression line,
               thick dotted line is the same regression line adjusted for errors in the distances, and 
               the solid line is the mass--radius relationship adopted for our new distance method, 
               see Eq.~(\ref{eq:constants3}).}
\label{fig:bulgemassaradie}
\end{figure}

Will the same distance scale work for both local and Bulge PNe? Present models of galactic 
evolution as well as observations point to that the progenitor stars are very different for PNe in the 
various components of the Milky Way. If the mass--radius relationships of the two different populations 
are similar, it indicates that the origin of the relations's appearance could be fundamental. Inseparable 
relationships therefore demonstrates that a common distance scale should work. 
Fig.~\ref{fig:bulgemassaradie} shows this distribution for our Bulge sample, derived from 
Eqs.~(\ref{eq:shklovsky4}) and (\ref{eq:pnsize}), assuming they all lie at a distance of 8.0 kpc. A linear
regression line for this distribution is also shown (thin dotted line), as well as the adopted 
mass--radius relationship from Sect.~\ref{sec:newmethod} (solid line). By assuming that all PNe lie at 
8.0 kpc we automatically introduce an error of 25\% in the distances. This has the effect (as previously 
discussed) of steepening the observed distribution. Reducing the regression coefficients of the of the 
Bulge sample, in the same way as in Sect.~\ref{sec:newmethod}, puts the Bulge mass-radius relation (strong 
dotted line) in closer agreement with our relationship for more local PNe.

Due to the likeness, although not perfect, between the mass--radius relations for the Bulge sample
and our calibration PNe, we believe that our new distance method is applicable to PNe in the Bulge as 
well.

\subsection{Efficiency of new Bulge criteria} \label{sec:efficiency}

\begin{figure}
      \resizebox{\hsize}{!}{\includegraphics{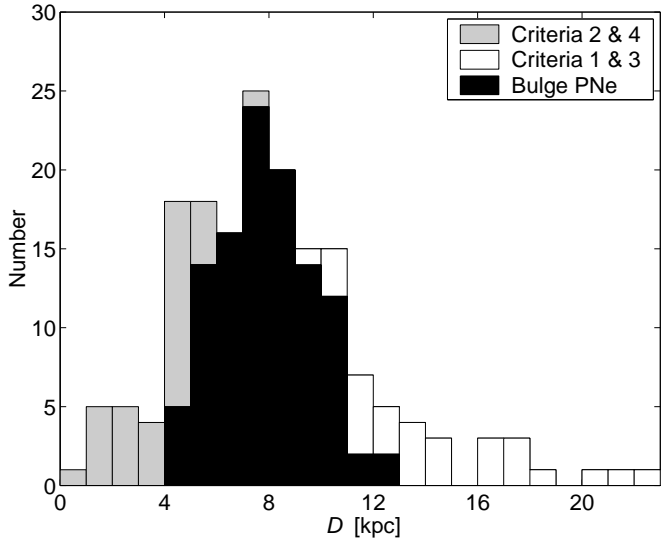}}
      \caption{The distribution of PNe towards the Bulge. White areas are background PNe that are rejected
               by criteria ``1'' and ``3'', and grey areas are foreground PNe sorted out by criteria ``2''
               and ``4'' (see  Fig.~\ref{fig:criteria} and Sect.~\ref{sec:sizelimit}). The black area is 
               the 109 PNe that remains after all criteria have been applied.}
\label{fig:effects}
\end{figure}
Fig.~\ref{fig:effects} shows how the distribution of the PNe towards the Bulge is influenced by our new 
criteria. Distances are computed by our new method. Starting with 174 PNe that only fulfill the coordinate
criteria, the foreground objects are reduced by criteria ``2'' and ``4''  (lines 2 and 4 in 
Fig.~\ref{fig:criteria}), and the background objects diminish through criteria ``1'' and ``3'' (lines 1 
and 3 in Fig.~\ref{fig:criteria}). Clearly our new criteria are not only efficient in reducing 
foreground objects, but it also takes care of objects beyond the Bulge.

Actually the number of background PNe appears to be somewhat high. Since Bulge nebulae are mostly 
seen at latitudes of 2 degrees or higher, this compares to $z=200$ pc on the near side of the Bulge and 
$z=350$ pc on the far side. According to Zijlstra \& Pottasch (\cite{zijlstra}) the scale height of PNe is
about 250 pc. This means that there will be fewer objects causing background confusion. A plausable 
explanation is that the criteria also exclude those PNe for which the radio fluxes are in doubt, that is, 
extended PNe with a low radio flux (see Fig.~\ref{fig:criteria}). This effect is as important as the 
removal of non--Bulge members, since it certainly will influence the interpretation of the derived
Bulge distributions.
 
Also a few PNe that, according to their distances, are members of the Bulge have been weeded out, as we 
suspected in Sect.~\ref{sec:criteria}. The probability of deleting a true member of the Bulge is anyway much 
smaller than deleting the outliers. It is important to bear in mind that the new criteria were set in 
order to achieve a sample free from PNe in front of and beyond the Bulge.

\subsection{New method vs. Old methods - using Bulge PNe} 
\label{sec:newdistances}

A way to test if a distance method is consistent is to try it on two different sets of PNe: one where they
are supposed to be optically thin, and one where they are supposed to be optically thick. The mass limit 
when a PN becomes optically thin is somewhat uncertain but should approximately be when the ionized 
envelope reaches a radius of $\sim0.2$ pc. For GBPNe, which are supposed to have an average distance of 
8 kpc, this radius compares to an angular diameter of $\sim5\arcsec$, which is by coincidence where our 
sample of Bulge nebulae peaks, see Fig.~\ref{fig:size_and_flux}. Splitting the sample in two, one with 
angular diameters larger than or equal to $4.6\arcsec$ and one with angular diameters smaller than 
$4.6\arcsec$ gives two samples consisting of 56 and 53 PNe respectively.

What kind of differences can we expect from a such division? If we go back to the classical Shklovsky 
method which assumes a constant mass (usually 0.2 M$_{\sun}$) for all PNe, both large and small, we might 
see how this method overestimates the masses for small PNe, and therefore also their distances. Larger 
PNe, on the other side, which actually seem to cluster around an ionized mass of $\sim0.2$ M$_{\sun}$, 
should have distances with a quite high degree of accuracy. If our new method, as well as the other 
previously discussed methods, is to be used as a good distance estimator for PNe, the difference between 
the distance distributions for the small and the large nebulae should preferably be small. It is important
to point out that it is not very likely that they should be identical, due to the difficulty of achieving 
a statistically complete non--biased sample of PNe in the Bulge.  

\begin{figure*}
      \resizebox{\hsize}{!}{\includegraphics{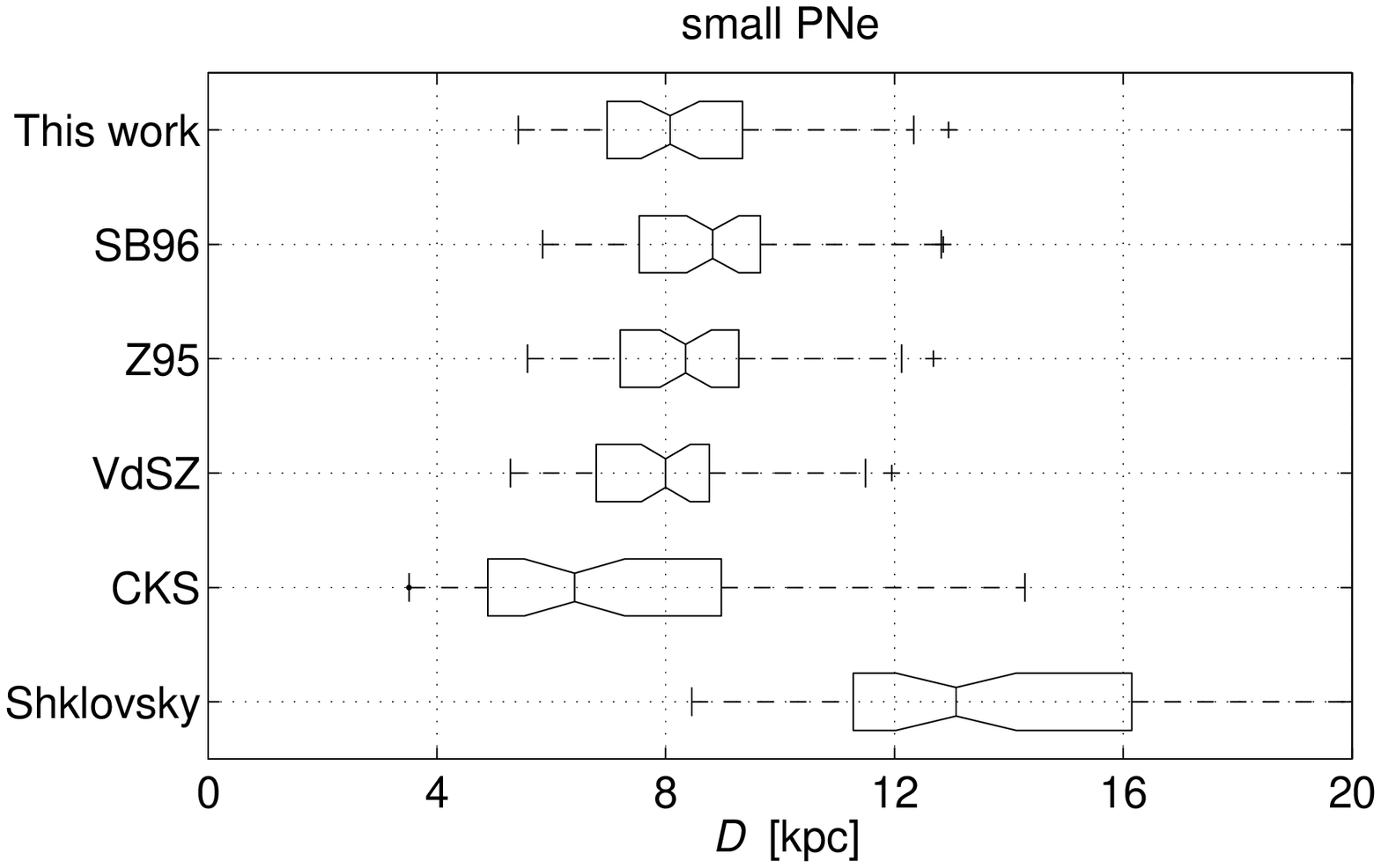}
                            \includegraphics{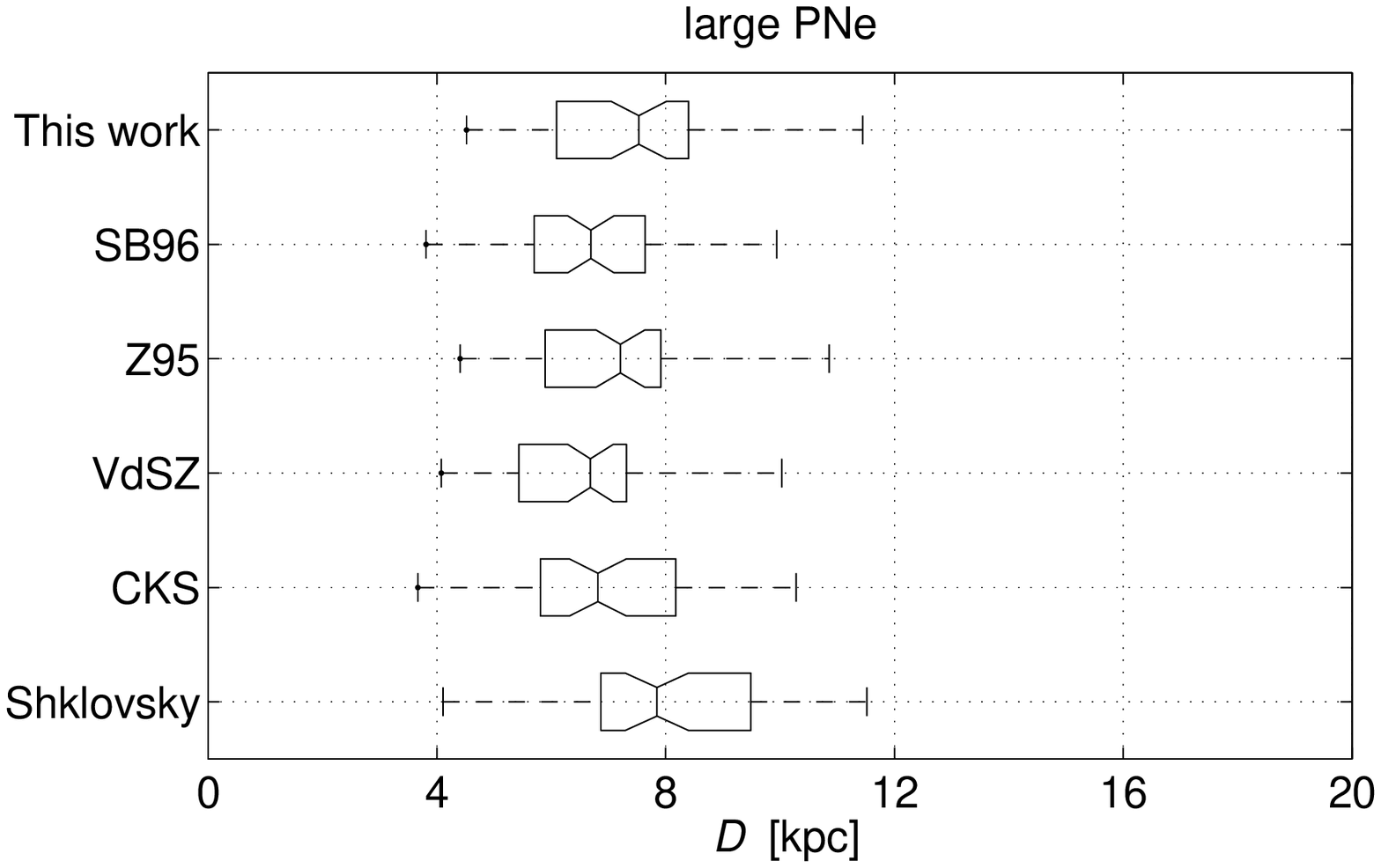}}
      \caption{Comparison of our new method with the old methods for PNe that fulfill the new Bulge 
               criteria. CKS, VdSZ, Z95 and SB96 distances were derived from Eqs.~(\ref{eq:cks}), 
               (\ref{eq:vdsz}), (\ref{eq:z95}) and (\ref{eq:sb96}), and the Shklovsky distances from 
               Eq.~(\ref{eq:shklovsky3}). The distributions are shown as box--plots where the boxes have 
               lines at the lower quartile, median and upper quartile values. The whiskers are lines 
               extending from each end of the box to show the extent of the rest of the data. Outliers are
               marked with plus signs and are defined as those PNe whose distance is more than 1.5 times 
               the interquartile range away from the top or bottom of the box. The notches in the box are 
               graphic confidence intervals about the median of the sample. The number of small PNe is 53,
               and large PNe is 56 in all samples.}
\label{fig:other2}
\end{figure*}

Fig.~\ref{fig:other2} shows the distance distributions of the 109 Bulge PNe, divided into the large and 
small samples, and Table~\ref{tab:statistics} shows some statistics.
\begin{table}[b]
 \centering
 \caption[]{Statistics for the distance distributions of the different methods for the Bulge sample. 
            $D_{0.5}$ is the median value, $\langle D \rangle$ the mean, and $\sigma_{D}$ the standard 
            deviation. `S' means small, `L' large, `A' all, and they are 53, 56 and 109 respectively.}
 \begin{tabular}{rccc}
 \hline\noalign{\smallskip}
                     &  $D_{0.5}$ [kpc] & $\langle D \rangle$ [kpc] &  $\sigma_{D}$ [kpc] \\
                     & A~~~~S~~~~L      & A~~~~S~~~~L               & A~~~~S~~~~L         \\
 \noalign{\smallskip}
 \hline\noalign{\smallskip}
 \noalign{\smallskip}
     This Work       & 7.8~~8.1~~7.5    & 7.9~~8.2~~7.6             & 1.8~~1.8~~1.8       \\
     SB96            & 7.6~~8.8~~6.7    & 7.7~~8.7~~6.6             & 1.9~~1.7~~1.4       \\
     Z95             & 7.8~~8.3~~7.2    & 7.8~~8.3~~7.2             & 1.8~~1.7~~1.7       \\
     VdSZ            & 7.2~~8.0~~6.7    & 7.3~~7.9~~6.7             & 1.7~~1.6~~1.5       \\
     CKS             & 6.6~~6.4~~6.8    & 7.0~~7.1~~6.9             & 2.1~~2.7~~1.5       \\
     Shklovsky       & 9.9~13.1~~7.8    & 11.1~13.8~8.6             & 4.8~~3.6~~4.4       \\
 \hline
 \end{tabular}
\label{tab:statistics}
\end{table}

\subsubsection{Large PNe} \label{sec:largepn}

Although one might expect the distribution of larger nebulae to peak at a somewhat shorter distance than 
the conventional 8 kpc to the GC, the large discrepancies for CKS, VdSZ, Z95 and SB96, where more than 
75\% of the samples are located on this side of the GC, are rather unexpected. Especially for VdSZ, that 
used GBPNe in their calibration, the result may be a consequence of the neglection of the background 
objects in their Bulge sample. Our method has on the other side a distribution that is more centered 
around the GC, although not perfectly. 

One reason to why the distributions tend to shorter distances is probably an effect of the 
non--completeness of the Bulge sample. The surface brightness of a PN, which is a distance independent 
parameter, is on average lower for a large nebula. Due to extinction in the Galaxy, PNe beyond the GC are 
more likely to fall below the limit of detectability, compared to those on this side. This will show as 
an increase of PNe on this side of the GC relatively PNe beyond the GC.

A second possible explanation, as mentioned in Sect.~\ref{sec:efficiency}, is that the Bulge criteria
also exclude those PNe for which the radio fluxes are in doubt. 

Another possible reason is that the large nebulae diameters are dominated by optical determinations, see 
Fig.~\ref{fig:size_and_flux}. As we saw in Sect.~\ref{sec:bulgesample} from high-resolution images of some 
of the PNe that have both optical and radio diameters, the ``correct'' values should be somewhere in 
between the radio and the optical. This means that optical values are on average slightly too large and 
the radio values on average too small. This has the effect of placing large PNe at too short distances and
small PNe at too long distances. How large these effects are on the distance distribution of the large PNe
is obscure, but we believe that the discrepancies that the old methods show are too large.

\subsubsection{Small PNe}

In practice all methods (except CKS and Shklovsky) show a distribution that is well centered around the GC
for the small PNe. A general trend is also that the small PNe give distance distributions that peaks at 
larger distances than for the large PNe. The causes for this are almost certainly of the same types as 
discussed in Sect.~\ref{sec:largepn}.

\subsubsection{Whole sample}

It is striking to see how well our method coincides with the method by Z95 when both small and large PNe 
are counted in. One could maybe draw the conclusion that this is an effect of the dominance of PNe from 
Zhang~(\cite{zhang}) in our calibration sample B. But if the calibration from Sect.~\ref{sec:newmethod} is 
repeated with only PNe from this sample, the regression coefficients, not corrected for the increased 
steepening, become
\begin{equation}
      \alpha=1.55\pm0.04, \,\,\,\,\,\,\,\,
      \beta=0.86\pm0.05.
\label{eq:constants2}
\end{equation}
i.e. quite different from the ones we achieved. Also, since these PNe were down--weighted by 0.5 with 
respect to those in sample A, we do not believe that our method is too similar to Z95 although the 
distribution (Table~\ref{tab:statistics}) indicate that they are. Instead we interpret the likeness 
between the two methods as a confirmation of the statistical methods as distance indicators, although they
differ in appearance, see Eqs.~(\ref{eq:z95}) and (\ref{eq:new2}).

Our method and Z95 are the ones that center nearest to 8.0 kpc. The other methods give distances 
considerably below this well-established distance to the GC. Since we do not know the completeness of the 
Bulge sample, it is not possible to say which of the methods that is the best. One thing there is no 
doubts about though, is that the CKS and the Shklovsky methods give distances that are far too short and 
too long respectively.

\subsection{The legality of statistical distances}

Table~\ref{tab:statistics} shows the spread of distances to bulge nebulae found using the different 
methods discussed. All recent investigations have values around 1.8 kpc. Assuming that the probable error 
in distance for one nebula is around 15\%, and that the Bulge has a radius of 1.5 kpc, we would in fact 
expect a spread of around 1.9 kpc. Since our strong membership criteria certainly exclude some true Bulge 
nebulae from our sample, the errors are probably somewhat larger, but most probably less than 30\% as also
indicated from our calibration sample (Sect.~\ref{sec:massradius}). 

This is comparable with errors expected from several ``individual'' distance determination methods. For 
example, Mendez et al.~(\cite{mendez2}) estimate that errors in distances from high--resolution 
spectroscopic determinations of effective temperatures and surface gravities of central stars of PNe 
amounts to $\sim25\%$. The method using resolved binary components also involves errors in absolute 
magnitude and reddening corrections of at least 0.3 mag. (Ciardullo et al.,~\cite{ciardullo}). This 
corresponds to errors in distance of around 15\%. A more realistic(?) estimate of a total error 
approaching 0.5 mag. correspond to a 25\% error in distance. Only a few objects have trigonometric 
parallaxes determined with errors smaller than 30\%. Other proposed methods have also comparable or higher
uncertainties.

From Fig.~\ref{fig:other2} and Table~\ref{tab:statistics} it is difficult to objectively decide which of
the discussed methods that give the best result for the Bulge test sample. The original Shklovsky method 
and the CKS relation are obviously inferior to the other methods. VdSZ fail to give the expected distance 
to the Bulge, while SB96 gives a relatively large difference between the two subsamples. Z95 and the 
present investigation give practically equally good results for the Bulge nebulae.  

Due to the uncertainty of the appearance of the mass--radius relationship for very compact and very
extended PNe, a warning may be in place to use the statistical methods for these objects. As can be seen 
from Fig.~\ref{fig:massaradie} the PNe within the central rectangle are well represented by a linear
function, and for those we expect our method to give distances with high accuracy. Leaving out extreme PNe
is not a large limitation on the usefulness of the method since the majority of PNe are in an evolutionary
stage such that they fall into the central group.

\section{Conclusions} \label{sec:conclusions}

We have recalibrated the correlation between the ionized mass and the nebular radius for Galactic PNe to 
enable an analyze of the distances that statistical methods produce. For the first time the 
mass--radius relationship has been compensated for the steepening and shift, which is the inevitable 
effect that errors in the distances have on the distribution of the nebulae in the `$\log M$--$\log R$' 
plane. In order to compare our new method with other statistical distance methods we reworked the 
criteria for Bulge membership. The old criteria did not pay attention to the problem of background 
objects, leading to Bulge samples that were polluted by PNe at long distances. Our new criteria are quite 
tight, meaning that they almost certainly also weed out true members of the Bulge. Since they were taken 
out in order to produce a sample that is as clean from outliers as possible, this is an unavoidable 
effect.

Applying the new criteria on PNe from the SECAT we derived a sample of 109 nebulae that are likely to be 
located within the Bulge. Using this sample, and calculating distances using our new method as well as the
old methods we show that our method gives distances with a high degree of accuracy, fully comparable to 
individual distance determinations. Also noticed is that all the older statistical methods give shorter 
distances.

We also divided the Bulge sample into two samples, one with optically thick PNe, and one with optically 
thin PNe. This was done in order to check on the consistency of the methods. All methods (except CKS) give
distance distributions at longer distances for the small (optically thick) PNe compared to the larger 
ones. Also all methods, except our, give distances to the large nebulae that are too short. We believe 
that this is partly due to the incompleteness of the Bulge sample. It is not likely, though, that it is 
completely responsible for the large discrepancy from the 8 kpc distance to the GC that the old methods 
show for large PNe. Small PNe show in general a well centered distance distribution around the GC.

With this work we believe that the method of determining distances to PNe in a statistical way has been 
settled. The long distance scales of VdSZ and Z95 are in a way confirmed, although small differences 
persist. The way VdSZ selected their Bulge sample, used for calibration, is a reason to why differences 
are found. Older, and shorter, distance scales, such as the one by Daub (recalibrated by CKS) are 
basically outdated.

\begin{acknowledgements}
We wish to thank our referee Dr.~Albert Zijlstra for his remarks which allowed us to see more clearly, 
improving the analysis and text of the paper. This investigation is partly based  on observations made 
with the NASA/ESA Hubble Space Telescope, obtained from the data archive at the Space Telescope Institute.
STScI is operated by the association of Universities for Research in Astronomy, Inc. under the NASA 
contract  NAS 5-26555.
\end{acknowledgements}


\onecolumn

\section*{Appendix}

\setcounter{table}{0}
\renewcommand{\thetable}{A}

\begin{table}[ht]
 \centering
 \caption[]{{\bf Sample A:} 41 PNe with at least three individual distance determinations. For 
            references of the adopted values on $D$, $\theta$ and $S_{\rm 5\,GHz}$, see 
            Table~\ref{tab:sample_C}.} 
 \begin{tabular}{clrrcrrll}
 \hline\noalign{\smallskip}
  PN G  &  ~~Name  &  $D$  [pc] & $N$ & $\Delta D / D$ & $\theta$  [\arcsec]  & $S_{\rm 5\,GHz}$  [mJy] &  $M_{\rm eff}$ [{\it M}$_{\sun}$]    & $R$ [pc]   \\
 \noalign{\smallskip}
 \hline\noalign{\smallskip}
 \noalign{\smallskip}
    002.4$+$05.8  &   \object{NGC 6369}   &     1490  &   4 & 0.15 &  35.0   &  1958~~~~~	& ~~~~0.61   & ~0.13    \\
    009.4$-$05.0  &   \object{NGC 6629}   &     1956  &   8 & 0.10 &  16.0   &   276~~~~~	& ~~~~0.14   & ~0.076   \\
    010.1$+$00.7  &   \object{NGC 6537}   &     2110  &   4 & 0.11 &  10.0   &   628~~~~~	& ~~~~0.13   & ~0.051   \\
    010.8$-$01.8  &   \object{NGC 6578}   &     2350  &   4 & 0.17 &   8.5   &   164~~~~~	& ~~~~0.066  & ~0.048   \\
    011.7$-$00.6  &   \object{NGC 6567}   &     1271  &  11 & 0.06 &   7.0   &   167~~~~~	& ~~~~0.011  & ~0.022   \\
    023.9$-$02.3  &   \object{M1-59}      &     1293  &   3 & 0.11 &   6.5   &   130~~~~~	& ~~~~0.0088 & ~0.020   \\
    029.2$-$05.9  &   \object{NGC 6751}   &     1961  &   4 & 0.19 &  22.0   &    62~~~~~	& ~~~~0.11   & ~0.10    \\
    033.1$-$06.3  &   \object{NGC 6772}   &     1260  &   4 & 0.02 &  65.0   &    88~~~~~	& ~~~~0.21   & ~0.20    \\
    033.8$-$02.6  &   \object{NGC 6741}   &     1477  &   6 & 0.13 &   8.0   &   194~~~~~	& ~~~~0.020  & ~0.029   \\
    036.1$-$57.1  &   \object{NGC 7293}   &      216  &   9 & 0.11 & 660.0   &  1292~~~~~	& ~~~~0.32   & ~0.35    \\
    037.7$-$34.5  &   \object{NGC 7009}   &     1076  &  10 & 0.17 &  28.5   &   736~~~~~	& ~~~~0.12   & ~0.074   \\
    041.8$-$02.9  &   \object{NGC 6781}   &     1567  &   3 & 0.02 & 108.0   &   340~~~~~	& ~~~~1.6    & ~0.41    \\
    043.1$+$37.7  &   \object{NGC 6210}   &     1630  &   4 & 0.13 &  30.0   &   258~~~~~	& ~~~~0.22   & ~0.12    \\
    045.4$-$02.7  &   \object{Vy 2-2}     &     3800  &   3 & 0.11 &   1.0   &   264~~~~~	& ~~~~0.011  & ~0.0092  \\
    045.7$-$04.5  &   \object{NGC 6804}   &     1525  &   4 & 0.12 &  35.0   &   136~~~~~	& ~~~~0.17   & ~0.13    \\
    046.4$-$04.1  &   \object{NGC 6803}   &     1733  &   6 & 0.18 &   5.5   &   109~~~~~	& ~~~~0.013  & ~0.023   \\
    069.4$-$02.6  &   \object{NGC 6894}   &     1428  &   5 & 0.11 &  42.0   &    61~~~~~	& ~~~~0.13   & ~0.15    \\
    084.9$-$03.4  &   \object{NGC 7027}   &     1045  &   8 & 0.13 &  14.0   &  6130~~~~~	& ~~~~0.11   & ~0.036   \\
    088.7$-$01.6  &   \object{NGC 7048}   &     1725  &   4 & 0.15 &  61.0   &    37~~~~~	& ~~~~0.28   & ~0.26    \\
    093.4$+$05.4  &   \object{NGC 7008}   &      883  &   6 & 0.14 &  86.0   &   217~~~~~	& ~~~~0.21   & ~0.18    \\
    118.8$-$74.7  &   \object{NGC 246}    &      496  &  10 & 0.05 & 245.0   &   262~~~~~	& ~~~~0.26   & ~0.29    \\
    197.8$+$17.3  &   \object{NGC 2392}   &     1176  &  10 & 0.19 &  19.5   &   262~~~~~	& ~~~~0.051  & ~0.056   \\
    205.1$+$14.2  &   \object{A21}        &      513  &   6 & 0.09 & 550.0   &   390~~~~~	& ~~~~1.2    & ~0.68    \\
    215.6$+$03.6  &   \object{NGC 2346}   &      904  &  14 & 0.09 &  52.0   &    86~~~~~	& ~~~~0.066  & ~0.11    \\
    234.8$+$02.4  &   \object{NGC 2440}   &     1753  &  12 & 0.14 &  20.0   &   397~~~~~	& ~~~~0.18   & ~0.085   \\
    243.3$-$01.0  &   \object{NGC 2452}   &     3217  &   7 & 0.08 &  19.0   &    58~~~~~	& ~~~~0.29   & ~0.15    \\ 
    261.0$+$32.0  &   \object{NGC 3242}   &      901  &  11 & 0.18 &  25.0   &   860~~~~~	& ~~~~0.069  & ~0.055   \\
    265.7$+$04.1  &   \object{NGC 2792}   &     2507  &  6  & 0.07 &  13.0   &   115~~~~~	& ~~~~0.12   & ~0.079   \\
    272.1$+$12.3  &   \object{NGC 3132}   &      644  &  9  & 0.08 &  30.0   &   228~~~~~	& ~~~~0.020  & ~0.047   \\
    278.1$-$05.9  &   \object{NGC 2867}   &     1908  &  7  & 0.14 &  14.0   &   274~~~~~	& ~~~~0.11   & ~0.065   \\
    286.3$-$04.8  &   \object{NGC 3211}   &     2833  &  4  & 0.14 &  16.0   &    86~~~~~	& ~~~~0.20   & ~0.11    \\
    294.1$+$43.6  &   \object{NGC 4361}   &     1442  &  6  & 0.11 &  63.0   &   218~~~~~	& ~~~~0.45   & ~0.22    \\
    294.6$+$04.7  &   \object{NGC 3918}   &     1537  &  6  & 0.15 &  19.0   &   856~~~~~	& ~~~~0.17   & ~0.071   \\
    298.3$-$04.8  &   \object{NGC 4071}   &     1133  &  3  & 0.16 &  63.0   &    26~~~~~	& ~~~~0.085  & ~0.17    \\
    307.2$-$03.4  &   \object{NGC 5189}   &     1266  &  5  & 0.15 & 140.0   &   462~~~~~	& ~~~~1.6    & ~0.43    \\
    309.1$-$04.3  &   \object{NGC 5315}   &     2564  &  5  & 0.14 &   6.0   &   444~~~~~	& ~~~~0.080  & ~0.037   \\
    316.1$+$08.4  &   \object{He2-108}    &     4900  &  3  & 0.15 &  11.0   &    40~~~~~	& ~~~~0.30   & ~0.13    \\
    320.1$-$09.6  &   \object{He2-138}    &     2775  &  4  & 0.11 &   7.0   &    84~~~~~	& ~~~~0.053  & ~0.047   \\
    341.8$+$05.4  &   \object{NGC 6153}   &     1668  &  4  & 0.03 &  24.0   &   550~~~~~	& ~~~~0.24   & ~0.097   \\
    350.9$+$04.4  &   \object{H2-1}       &     4167  &  3  & 0.07 &   2.8   &    64~~~~~	& ~~~~0.032  & ~0.028   \\
    358.9$-$00.7  &   \object{M1-26}      &     1733  &  6  & 0.11 &   4.8   &   389~~~~~	& ~~~~0.020  & ~0.020   \\
 \hline
 \end{tabular}
 \label{tab:sample_A}
\end{table}

\renewcommand{\thetable}{B}
\begin{table}
 \centering
 \caption[]{{\bf Sample B:} 32 PNe with only two individual distance determinations. For 
            references of the adopted values on $D$, $\theta$ and $S_{\rm 5\,GHz}$, see 
            Table~\ref{tab:sample_C}.}
 \begin{tabular}{clrrcrrll}
 \hline\noalign{\smallskip}
  PN G & ~~Name & $D$  [pc] & $N$ & $\Delta D / D$ & $\theta$  [\arcsec]  & $S_{\rm 5\,GHz}$  [mJy] & $M_{\rm eff}$  [{\it M}$_{\sun}$]   & $R$  [pc]   \\

 \noalign{\smallskip}
 \hline\noalign{\smallskip}
 \noalign{\smallskip}
    000.3$+$12.2  &   \object{IC 4634}    &     4900  & 2 & 0.10      &    8.4      &   123~~~~~	  & ~~~~0.35   & ~0.10   \\
    002.0$-$13.4  &   \object{IC 4776}    &     7350  & 2 & 0.14      &    7.5      &    69~~~~~	  & ~~~~0.61   & ~0.13   \\
    002.1$-$04.2  &   \object{H 1-54}     &    11650  & 2 & 0.02      &    2.0      &    26~~~~~	  & ~~~~0.16   & ~0.057  \\
    002.4$-$03.7  &   \object{M 1-38}     &     6400  & 2 & 0.05      &    3.3      &    34~~~~~	  & ~~~~0.088  & ~0.051  \\
    003.7$-$04.6  &   \object{M 2-30}     &    12100  & 2 & 0.15      &    9.0      &    14~~~~~	  & ~~~~1.3    & ~0.26   \\
    005.8$-$06.1  &   \object{NGC 6620}   &     4400  & 2 & 0.02      &    8.0      &    16~~~~~	  & ~~~~0.090  & ~0.085  \\
    008.0$+$03.9  &   \object{NGC 6445}   &     2250  & 2 & 0.11      &   33.0      &   364~~~~~	  & ~~~~0.67   & ~0.18   \\
    008.3$-$07.3  &   \object{NGC 6644}   &     3000  & 2 & 0.07      &    2.6      &    98~~~~~	  & ~~~~0.016  & ~0.019  \\
    025.8$-$17.9  &   \object{NGC 6818}   &     2230  & 2 & 0.01      &   30.0      &   291~~~~~	  & ~~~~0.51   & ~0.16   \\
    065.0$-$27.3  &   \object{Ps 1}       &    11550  & 2 & 0.17      &    2.4      &     5~~~~~	  & ~~~~0.092  & ~0.067  \\
    074.5$+$02.1  &   \object{NGC 6881}   &     6250  & 2 & 0.02      &    2.6      &   121~~~~~	  & ~~~~0.11   & ~0.039  \\
    093.3$-$02.4  &   \object{M 1-79}     &     2350  & 2 & 0.15      &   33.0      &    19~~~~~	  & ~~~~0.17   & ~0.19   \\
    107.6$-$13.3  &   \object{Vy 2- 3}    &    11350  & 2 & 0.07      &    4.5      &     3~~~~~	  & ~~~~0.18   & ~0.12   \\
    111.8$-$02.8  &   \object{Hb 12}      &     5750  & 2 & 0.15      &    1.0      &   367~~~~~	  & ~~~~0.037  & ~0.014  \\
    118.0$-$08.6  &   \object{Vy 1- 1}    &     6650  & 2 & 0.04      &    6.0      &    20~~~~~	  & ~~~~0.18   & ~0.097  \\
    161.2$-$14.8  &   \object{IC 2003}    &     9750  & 2 & 0.04      &    9.0      &    47~~~~~	  & ~~~~1.3    & ~0.21   \\
    190.3$-$17.7  &   \object{J 320}      &     6300  & 2 & 0.13      &    9.0      &    28~~~~~	  & ~~~~0.35   & ~0.14   \\
    238.0$+$34.8  &   \object{A 33}       &     1210  & 2 & 0.04      &  280.0      &    39~~~~~	  & ~~~~1.1    & ~0.82   \\
    291.6$-$04.8  &   \object{IC 2621}    &     4750  & 2 & 0.01      &    5.0      &   195~~~~~	  & ~~~~0.19   & ~0.058  \\
    296.3$-$03.0  &   \object{He 2- 73}   &     6500  & 2 & 0.03      &    4.0      &    72~~~~~	  & ~~~~0.18   & ~0.063  \\
    312.3$+$10.5  &   \object{NGC 5307}   &     2470  & 2 & 0.05      &   16.0      &    92~~~~~	  & ~~~~0.14   & ~0.096  \\
    321.3$+$02.8  &   \object{He 2-115}   &     4000  & 2 & 0.05      &    3.0      &   132~~~~~	  & ~~~~0.047  & ~0.029  \\
    323.9$+$02.4  &   \object{He 2-123}   &     6150  & 2 & 0.01      &    4.6      &    84~~~~~	  & ~~~~0.21   & ~0.069  \\
    325.8$-$12.8  &   \object{He2-182}    &     5350  & 2 & 0.18      &    2.5      &    66~~~~~          & ~~~~0.052  & ~0.032  \\
    326.0$-$06.5  &   \object{He2-151}    &     6800  & 2 & 0.01      &    1.0      &     7~~~~~          & ~~~~0.0078 & ~0.017  \\
    327.1$-$02.2  &   \object{He 2-142}   &     3850  & 2 & 0.04      &    3.6      &    70~~~~~	  & ~~~~0.041  & ~0.034  \\
    329.0$+$01.9  &   \object{Sp 1}       &     1450  & 2 & 0.10      &   72.0      &    76~~~~~	  & ~~~~0.33   & ~0.25   \\
    331.3$+$16.8  &   \object{NGC 5873}   &     9200  & 2 & 0.18      &    7.0      &    45~~~~~	  & ~~~~0.78   & ~0.16   \\
    331.4$-$03.5  &   \object{He2-162}    &     4000  & 2 & 0.08      &   5.0       &    23~~~~~          & ~~~~0.042  & ~0.049  \\
    355.1$-$06.9  &   \object{M 3-21}     &     8500  & 2 & 0.07      &    5.0      &    28~~~~~	  & ~~~~0.30   & ~0.10   \\
    355.7$-$03.5  &   \object{H 1-35}     &     8050  & 2 & 0.07      &    1.7      &    93~~~~~	  & ~~~~0.096  & ~0.033  \\
    355.9$-$04.2  &   \object{M 1-30}     &     7950  & 2 & 0.01      &    3.5      &    31~~~~~	  & ~~~~0.16   & ~0.067  \\

 \hline
 \end{tabular}
 \label{tab:sample_B}
\end{table}

\renewcommand{\thetable}{C}
\begin{table}
 \centering
 \caption[]{References for the distances ($D$), radio fluxes ($S_{\rm 5\,GHz}$) and the diameters ($\theta$) 
            in Tables~\ref{tab:sample_A} and \ref{tab:sample_B}.
 
 {\bf Distances:} \\ 
      (1) Acker (\cite{acker2}),
      (2) Acker et al. (\cite{acker}),
      (3) Bond and Ciardullo (\cite{bond}),
      (4) B\"assgen et al. (\cite{bassgen}),
      (5) Christianto and Seaquist (\cite{christianto}),
      (6) Ciardullo et al. (\cite{ciardullo}),
      (7) Costero et al. (\cite{costero}),
      (8) Gathier et al. (\cite{gathier}),
      (9) Gathier et al. (\cite{gathier2}),
      (10) Gurzadyan (\cite{gurzadyan})
      (11) Guti\'errez-Moreno et al. (\cite{gutierrez}),
      (12) Hajian et al. (\cite{hajian4}),
      (13) Hajian et al. (\cite{hajian3}),
      (14) Hajian et al. (\cite{hajian2}),
      (15) Harrington and Dahn (\cite{harrington}),
      (16) Harris et al. (\cite{harris}),
      (17) Huemer and Weinberger (\cite{huemer}),
      (18) Jacoby (\cite{jacoby}),
      (19) Kaler and Lutz (\cite{kaler}),
      (20) Kaler et al. (\cite{kaler2}),
      (21) Kohoutek (\cite{kohoutek}),
      (22) Lutz (\cite{lutz}),
      (23) Maciel (\cite{maciel5}),
      (24) Maciel and Cazetta (\cite{maciel}),
      (25) Maciel and Pottasch (\cite{maciel4}),
      (26) Martin (\cite{martin}),
      (27) Masson (\cite{masson3}),
      (28) Masson (\cite{masson}),
      (29) Meatheringham et al. (\cite{meatheringham}),
      (30) Mendez and Niemela (\cite{niemela}),
      (31) Mendez et al. (\cite{mendez}),
      (32) Napiwotzki and Sch\"onberner (\cite{napiwotzki}),
      (33) Pier et al. (\cite{pier}),
      (34) Pollacco and Ramsay (\cite{pollacco}),
      (35) Pottasch (\cite{pottasch3}),
      (36) Pottasch (\cite{pottasch6}),
      (37) Pottasch (\cite{pottasch5}),
      (38) Pottasch (\cite{pottasch10}),
      (39) Pottasch (\cite{pottasch}),
      (40) Pottasch et al. (\cite{pottasch2}),
      (41) Sabbadin (\cite{sabbadin}),
      (42) Saurer (\cite{saurer2}),
      (43) Saurer (\cite{saurer3}),
      (44) Zhang (\cite{zhang}),
      (45) Walsh et al. (\cite{walsh}). \\
 {\bf Fluxes:} \\
     (46)  Aaquist and Kwok (\cite{aakw}),
     (47)  Basart and Daub (\cite{basart}),
     (48)  Birkinshaw et al. (\cite{bshaw}),
     (49)  Gathier et al. (\cite{gathier5}),
     (50)  Gathier et al. (\cite{gathier4}),
     (51)  Higgs (\cite{higgs}),
     (52)  Isaacman (\cite{isaacman}),
     (53)  Milne and Aller (\cite{milne}), 
     (54)  Milne and Aller (\cite{milne3}),
     (55)  Phillips and Mampaso (\cite{phillips}),
     (56)  Pottasch and Zijlstra (\cite{pottasch9}),
     (57)  Purton et al. (\cite{purton}),
     (58)  Ratag and Pottasch (\cite{ratag}),
     (59)  Zijlstra et al. (\cite{zijlstra4}). \\
 {\bf Diameters:} \\ 
     (60)  Aaquist and Kwok (\cite{aakw}),
     (61)  Bedding and Zijlstra (\cite{bedding}),
     (62)  Cahn and Kaler (\cite{cahn2}),
     (63)  Chu et al. (\cite{chu}),
     (64)  Dopita et al. (\cite{dopita}),
     (65)  Gathier et al. (\cite{gathier4}),
     (66)  HST Archive broad band image,
     (67)  HST Archive H-alpha image,
     (68)  HST Archive [NII] image,
     (69)  Hua and Kwok (\cite{hua}),
     (70)  Mendez et al. (\cite{mendez2}),
     (71)  Milne and Aller (\cite{milne}),
     (72)  Moreno et al. (\cite{moreno}),
     (73)  Perek and Kohoutek (\cite{perek}),
     (74)  Seaquist and Davies (\cite{seaquist}),
     (75)  Zijlstra et al. (\cite{zijlstra4}).
 }

 \begin{tabular}{clll|clll}
 \hline\noalign{\smallskip}
  PN G              & $D$-ref                         & $S$-ref      & $\theta$-ref & PN G             & $D$-ref       & $S$-ref         & $\theta$-ref  \\

 \noalign{\smallskip}
 \hline\noalign{\smallskip}
  \textbf{Sample A} &                                 &              &              & 320.1$-$09.6     & 24,31,44      & 53,54           & 62,67    \\
  002.4$+$05.8      & 9,29,36,44                      & 53,54        & 63,66        & 341.8$+$05.4     & 29,36,41,44   & 53,54           & 62,63    \\
  009.4$-$05.0      & 9,14,24,29,31,36,41,44          & 53,54        & 63,66        & 350.9$+$04.4     & 24,31,44      & 53,54,59        & 67       \\
  010.1$+$00.7      & 1,9,29,36                       & 52,53,54,55  & 67           & 358.9$-$00.7     & 1,24,31,41,44 & 46,53,54        & 62,67    \\ 
  010.8$-$01.8      & 9,36,41,44                      & 53,54        & 63,66        &                  &               &                 &          \\
  011.7$-$00.6      & 1,8,9,25,35,36,37,41,44         & 53,54        & 63           &\textbf{Sample B} &               &                 &          \\
  023.9$-$02.3      & 1,35                            & 52,53,54,59  & 67           & 000.3$+$12.2     & 44            & 46,53,54,55,58  & 62,67    \\
  029.2$-$05.9      & 1,14,36,44                      & 53,54        & 63,68        & 002.0$-$13.4     & 44            & 53,54,55        & 62,63    \\ 
  033.1$-$06.3      & 1,29,36,41                      & 53,54,59     & 62           & 002.1$-$04.2     & 44            & 50,54           & 64,67    \\ 
  033.8$-$02.6      & 1,19,36,41,44                   & 52,53,54,59  & 63,68        & 002.4$-$03.7     & 44            & 50,54           & 65,73    \\ 
  036.1$-$57.1      & 15,16,24,31,35,37,39,44         & 53           & 71,75        & 003.7$-$04.6     & 44            & 54,59           & 72       \\ 
  037.7$-$34.5      & 1,20,23,24,25,29,31,35,36,44    & 53,54        & 63,68        & 005.8$-$06.1     & 44            & 53,54,55,56     & 62,72    \\ 
  041.8$-$02.9      & 1,36,41                         & 53,54,59     & 63,66        & 008.0$+$03.9     & 36,41         & 53,54           & 62       \\ 
  043.1$+$37.7      & 13,25,44                        & 53,54        & 63,67        & 008.3$-$07.3     & 25,44         & 50.55           & 62,63    \\ 
  045.4$-$02.7      & 5,44                            & 57           & 67,74        & 025.8$-$17.9     & 10,29         & 53,54           & 67       \\ 
  045.7$-$04.5      & 1,36,41                         & 53,54        & 63,66        & 065.0$-$27.3     & 25,41         & 48,49,53,54     & 67       \\ 
  046.4$-$04.1      & 1,9,35,36,41                    & 53,54,55     & 62,63        & 074.5$+$02.1     & 44            & 46,55           & 60,67    \\ 
  069.4$-$02.6      & 1,19,22,36,41                   & 59           & 63,66        & 093.3$-$02.4     & 42,43         & 59              & 62       \\ 
  084.9$-$03.4      & 12,18,27,28,37,39,40,41         & 59           & 63,67        & 107.6$-$13.3     & 44            & 59              & 62,67,75 \\ 
  088.7$-$01.6      & 1,17,41                         & 59           & 63           & 111.8$-$02.8     & 44            & 57              & 60,67    \\ 
  093.4$+$05.4      & 1,6,35,36,41                    & 59           & 63,66        & 118.0$-$08.6     & 44            & 46,55           & 60       \\ 
  118.8$-$74.7      & 2,3,35,36,39,45                 & 53,54        & 63           & 161.2$-$14.8     & 44            & 46,55           & 60,62,63 \\ 
  197.8$+$17.3      & 10,13,23,24,25,31,35,36,41,44   & 53,54        & 63,67        & 190.3$-$17.7     & 44            & 46,53,54,55     & 67       \\ 
  205.1$+$14.2      & 11,16,32,33,39                  & 53,54        & 69           & 238.0$+$34.8     & 6,39          & 53,54           & 69       \\  
  215.6$+$03.6      & 1,7,8,21,29,30,35,36,37,39      & 53           & 62,67        & 291.6$-$04.8     & 44            & 53,54           & 63       \\ 
  234.8$+$02.4      & 1,4,8,9,10,29,34,36,37,38,44    & 53,54,59     & 66,75        & 296.3$-$03.0     & 44            & 53,54           & 62       \\
  243.3$-$01.0      & 1,8,29,35,36,37,44              & 53,54        & 62,66        & 312.3$+$10.5     & 29,44         & 53,54           & 63,66    \\ 
  261.0$+$32.0      & 10,13,20,24,25,31,35,36,41,44   & 53,54        & 63,67        & 321.3$+$02.8     & 44            & 53,54           & 62,67    \\ 
  265.7$+$04.1      & 1,8,29,36,37,44                 & 53,54        & 63,66        & 323.9$+$02.4     & 44            & 53,54           & 62       \\
  272.1$+$12.3      & 6,8,29,35,36,39,41              & 53,54        & 63,67        & 325.8$-$12.8     & 24,31         & 53,54           & 67       \\   
  278.1$-$05.9      & 1,14,29,36,41,44                & 53,54        & 62,66        & 326.0$-$06.5     & 24,31         & 54              & 67       \\    
  286.3$-$04.8      & 8,36,29,44                      & 53,54        & 62           & 327.1$-$02.2     & 44            & 53,54           & 62,67    \\     
  294.1$+$43.6      & 24,29,31,37,39,44               & 53,54        & 63           & 329.0$+$01.9     & 1,41          & 53,54           & 62       \\       
  294.6$+$04.7      & 1,8,29,36,41,44                 & 53,54        & 62,66        & 331.3$+$16.8     & 44            & 53,54,55        & 62       \\        
  298.3$-$04.8      & 1,41,44                         & 53,54        & 62           & 331.4$-$03.5     & 44            & 53,54           & 70       \\  
  307.2$-$03.4      & 1,8,36,41                       & 53,54        & 62,66        & 355.1$-$06.9     & 44            & 53,54           & 73       \\
  309.1$-$04.3      & 1,8,36,44                       & 53,54        & 62,66        & 355.7$-$03.5     & 44            & 53,54,59        & 61       \\        
  316.1$+$08.4      & 24,31,44                        & 53,54        & 62           & 355.9$-$04.2     & 44            & 50              & 65       \\            
                                                     
 \hline
 \end{tabular}
 \label{tab:sample_C}
\end{table}

\end{document}